\documentstyle[aps,epsf,pra,preprint]{revtex}    
\begin{document}
\draft

%  Defining Macros 
\newcommand {\rv}{\vec{r}}
\def\vr{\vec{r}}
\def\wxr{ W_x (\vec{r})} 
\def\exr{ {\vec {\cal{ E}}}_{x}(\vec{r})} 
\def\ex{ \vec {\cal {E}}_{x}} 
\def\exc{ \vec {\cal {E}}_{xc}} 
\def\roex{ {\rho}_{x}(\vec{r},\vec{r'})} 
\newcommand {\rvp}{\vec{r'}}
\newcommand {\vrp}{\vec{r'}}
\newcommand {\vp}{\vec{p}}
\newcommand {\vpp}{\vec{p'}}
\newcommand {\gp}{\gamma{(\vp)}}
\newcommand {\jq}{J(q)}
\newcommand {\dmmom}{ {\Gamma}^{(1)}_{mom}( \vp \vert \vpp )}
\newcommand {\phippn}{ \Phi ( {\vec{p}}_1, {\vec{p}}_2, \ldots, {\vec{p}}_N)}
\newcommand {\phisp}{ \Phi^* (\vp,{\vec{p}}_2, \ldots, \vec{p}_N)}
\newcommand {\phipp}{ \Phi (\vpp,\vec{p}_2, \ldots, \vec{p}_N)}
\newcommand {\grr}{ {\Gamma}^{(1)} (\vec{r},\vec{r})} 
\def\spinj{ \sum_{j \atop {{\rm spin}\, i= {\rm spin}\, j}}}

\title{Atomic Compton Profiles within different exchange-only theories}

\author{Rajendra R. Zope}

\address{Department of Physics, University of Pune, Pune 411 007, Maharashtra,  India}

\author{Manoj K. Harbola}

\address{Laser Programme, Centre for Advanced Technology, Indore-452
013, Madhya Pradesh, India}

\author{Rajeev K. Pathak\footnote{Author to whom all correspondence may
be kindly addressed.}}

\address{Department of Physics, University of Pune, Pune 411 007, Maharashtra, India}

\date{\today}
\maketitle

\begin{abstract}

      The Impulse Compton Profiles (CP's) $J(q)$ and the
 $\langle\!{p^n}\rangle -$
expectation values for  some inert gas atoms(He-Kr) 
are computed and compared within the Harbola-Sahni (HS), Hartree-Fock (HF)
theories  and a Self Interaction Corrected (SIC) density functional model. 
The Compton profiles for excited states of Helium atom are also 
calculated.  While the calculated CP's are found to generally agree,
they  differ slightly from one another
for small values of the Compton parameter $q$ and are
  in good  agreement for large $q$ values.
The $<\!p^n\!>$ expectation values  within the three theories
are also found to be comparable.
 The HS formalism is seen to mimic HF reasonably well 
in the momentum space, establishing the logical consistency of the
former.

\end{abstract}

\vspace{0.8cm}

\narrowtext
            In the phenomenon of Compton scattering, the Compton cross section 
(of high energy X-ray or $\gamma$-ray photon inelastically scattered by 
electrons in matter) has a direct bearing on the electron-momentum density.
In the so-termed impulse approximation \cite{BW} the Compton cross section is 
proportional to an experimentally observable quantity, $viz.$  the Compton 
profile(CP), related to the electron momentum density, vide:
\begin{equation}
  J(q) =\int_{-\infty}^{\infty} \int_{-\infty}^{\infty} 
                   \gamma (p_x,p_y,q) dp_x dp_y .  
\end{equation} 
The momentum density  $\gp$ is  the diagonal $(i.e. ~\vpp = \vp )$ 
part of the full, reduced first
 order momentum space density matrix $\dmmom$ connected in turn, 
to the many-electron momentum-space wave function \cite{Mc_L} ~$\phippn$
 by 
\begin{eqnarray} 
& &  \dmmom     \nonumber \\ &=&  N \int \phisp \phipp d^3p_2 \ldots d^3p_N ,
	\end{eqnarray}
where a sum over spin may also be included.

	    Within the independent-electron approximation such as the
          Hartree-Fock (HF) theory, $\dmmom$  takes the form  \cite{SRG}
 
          $$  \dmmom  = \sum_i f_i  \phi^*_i (\vp) \phi_i (\vpp),$$
          with  $i$ sweeping through  the ``occupied'' 
          states with the occupancies $f_i$ and  $\phi_i(\vp)$, the
          momentum-space orbital, being the
          Fourier-transform of the coordinate-space orbital
          $\psi_i(\rv)$ related through 
	  (Hartree atomic units employed throughout):

          \begin{equation}
          \phi_i (\vp) = \frac{1}{(2\pi)^{3/2}}  \int e^{i\vp \cdot \rv} 
                    \psi_i(\rv) d^3r.
	  \end{equation}
     The  $<\!p^n\!>$- moments are defined in terms of  electron momentum 
     density (EMD) distribution by:
          \begin{equation}
             <\!p^n\!> = 4\pi \int_0^{\infty} p^{n+2} ~\gamma(p) dp,
     ~~~~~-2\le{n}\le 4,
	  \end{equation}
   where $\gamma(p) = \frac{1}{4\pi} \int \gamma(\vp)
          d\Omega_{{\hat{p}}}$ is the spherically averaged EMD, in turn
leading to the spherically averaged impulse Compton profile 
$J(q) = 2\pi \int_{\mid q \mid}^{\infty} \gamma (p) ~p ~dp $.
   These  $<\!p^n\!>$ expectation values sample the interior as well as exterior
regions of the EMD and are also related to atomic properties.  The
$<\!1/p\!>$ moment  is twice the peak value of the
impulse profile $J(0)$;  the $<\!p^2\!>$ moment is twice  the kinetic
energy ($=-E_{total}$, by the virial theorem)  while the $<\!p\!>$ moment 
is empirically found to be almost proportional to the exact 
Hartree-Fock  exchange energy \cite{RKP}.

      The recent Harbola-Sahni approach\cite{HS_1,HS_2} to the atomic structure
calculations proffers an attractive alternative to the conventional 
Hartree-Fock description.  The HS approach has been proven to be
successful in giving the total energies\cite{HS_9,HZP}
and co-ordinate space  properties \cite{Sen_1,Sen_2} practically 
of Hartree-Fock quality. %(with marginal differences, typically \%). 
In addition, coupled with local correlation, it also describes
the excited states of atoms quite accurately\cite{Deb}.
This success of  the HS formalism prompts one for its critical appraisal
in the momentum space through Compton profiles  and 
and the $<\!p^n\!>$ expectation values. We also  compare
these with the corresponding quantities calculated within the  
HF, the HS and the 
self interaction corrected (SIC) local density functional theories.
This study  is aimed at bringing out  how these $\vp -$space quantities
calculated using  the {\it local} and {\it orbital-independent}
prescription of HS compare with those of the HF theory, which
employs a nonlocal potential (in its {\it exact} exchange description),
and of SIC theory in which the effective
potential, although local, turns out to be orbital-dependent.
	In the following we first briefly describe the HF, HS and SIC theories
to highlight the differences among them.

	All the three theories HF, HS and SIC are independent particle
theories in which the electron orbitals are obtained by solving the
equations
(Hartree atomic units i.e. $\hbar = ~{\mid}{e}{\mid} = m =1 $ are used 
throughout herein), viz.
\begin{equation}  
\Bigl [- \frac{\nabla^2}{2} +  v_{H} (\vr)   
 + v_x (\vr) \Bigr ] \psi_i = \epsilon_i \psi_i ; ~~~i=1,2,....N ,
\end{equation}  
where $v_H(\vr) = v_{nuclear}(\vr) + \int \frac {\rho(\vrp)}{ \mid \vr - \vrp \mid } d^3r'$ 
is the Hartree potential and $v_x$ the exchange potential. Here
$\rho(\vr)$ is the electronic density given in terms of orbitals
$\psi_i(\vr)$ as $\rho(\vr) = \sum_i f_i \vert \psi_i (\vr) \vert^2$. The
differences in HF, HS and SIC precisely lies in the manner in which the 
 exchange potential is
prescribed in them. In HF which is the exact theory at the 
``exchange-only'' level, 
as noted above, the potential $v_x$ is 
{\it nonlocal} with its action on $\psi_i (\vr)$ is given by 
\begin{equation}  
 v_x(\vr) \psi_i (\vr) = \spinj  \int \frac {\psi_j^* (\vrp)
 \psi_j(\vr) \psi_i (\vrp)}{ \vert \vr - \vrp \vert} d^3r'.
\end{equation}

	On  the other hand, the exchange potential in the exchange-only HS
theory is local and is prescribed  as the work done in moving an electron in
the field of its Fermi hole\cite{HS_1}:
\begin{equation}  
 v_x (\vr) =  \wxr = - \int_{\infty}^{\vec{r}}  \ex \cdot d\vec{l}, 
\end{equation}  
where 
\begin{equation}  
\exr = \int \frac {\roex}{ \mid \vr - \vrp {\mid}^3} ~(\vr- \vrp) ~d^3 r' , 
          \label{Ex:eq}
\end{equation}  
is the exchange ``electric field'' due to the Fermi hole 
(or the ``Exchange hole'') $\rho_x(\vr,\vrp)$.

	In the SIC theory the exchange potential is calculated within
the local-density approximation (LDA) which is then {\em ad-hoc}ly  corrected 
for its self-interaction on an orbital-by-orbital basis\cite{SIC}. Thus the {\it orbital
dependent} SIC exchange potential is given as

\begin{equation}
v^i_{SIC,x}(\vr) = (\frac{-3}{4}) ({\frac{3}{\pi}})^{1/3} \rho^{1/3}(\vr) - 
\Bigl
\{ \int \frac{ \mid \psi_i (\vrp) \mid ^2}{ \mid \vr - \vrp \mid} d^3r'
+ (\frac{-3}{4}) ( \frac{6 \rho_i(\vr)}{\pi})^{1/3} \Bigr \}
\end{equation}  
where $\rho_i(\vr) = \mid \psi_i (\vr) \mid^2 $ is the orbital density.

%-------------------------------

	In the context of HS potential, it may be noted that
as recently established by Holas and March \cite{HM}, 
the Harbola-Sahni exchange-correlation potential can also be calculated  from the exact
second-order density matrix by employing the differential virial theorem. 
The results are consistent with, and provide the mathematical proof of the
formalism proposed by HS.  In addition, it also spells out how the kinetic
energy term missing \cite{HS_1,HS_2} from the HS
potential arises from the differences in the exact kinetic energy density  tensor
and its Slater-Kohn-Sham orbitals counterpart\cite{HM}.  Similar analysis carried
out   within the Hartree-Fock theory \cite{Sahni} reveal
that the difference between the exact KS exchange-only potential and
the HS potential is traced back to the differences in the kinetic energy 
density tensors of the 
HF theory and its local counterpart (as such, this
difference is indeed only marginal)\cite{Sahni}. Further, within a local
prescription, it is not clear as to how one incorporates the
kinetic-energy effects  directly in a self-consistent-field (SCF) scheme.

 In this work the spherically averaged Compton Profiles $J (q)$ 
within the HF theory are computed 
using  the near Hartree-Fock(NHF) quality wavefunctions that employ the STO
(Slater-Type-orbital) bases tabulated by Clementi and Roetti\cite{CR} and the values
of the $<\!p^n\!>$ moments are from Ref.\cite{Vega}.
  On the other hand, the orbitals with effective potential $W_x$
       are obtained  by a modified Herman-Skillman code\cite{Herm}.
             The calculated Compton profiles along with their 
available experimental \cite{ER,Tong_L} and accurate theoretical counterparts 
\cite{CP_CI} are tabulated in tables I-IV
for inert atomic systems He-Kr 
 while  the moments are  displayed
           in tables V-VII. 
        The Helium atom in its ground state has a single orbital,
        hence all  the CP's within these three exchange-only theories 
      practically coincide (cf. Table I).  
%--------------------
       For Ne, Ar and Kr,
        it is evident from the Tables II-V that for   $low ~q-$ values 
        the CP's differ from each
       other appreciably. For higher $q-$ values these theoretical
       (HF, HS, SIC) profiles are in  better agreement with one another. 
        In the low $q-$ region ($q < 0.5au),$ the SIC profiles are 
        seen to be the largest and the HS profiles are the smallest in 
        magnitude among the three theories. Beyond $q\sim 0.5 a.u.$
        the three profiles cross each other and are in good agreement
        in the asymptotic region. 
%
% Addition: revised 1
%
It is observed further that the ``experimental'' $J(q)$ is fairly well
estimated by the ``exchange only'' theories. 
         It is to be noted that an accurate theoretical $J(q)$ computation
  beyond  HF, viz. the configuration interaction calculation for $Ne$, 
  due to Tripathi {\it et al.}  \cite{CP_CI} while improving upon the HF-CP 
  still slightly underestimates the experimental $J(0)$ but overestimates
the intermediate  profile.
%
% Addition: revised 1 over
%
The higher values of $J(q)$ 
        in the SIC theory indicate that the momentum density is  localized
        near  the origin $\vp= \vec{0}$  in the SIC formalism. This  can  
also be seen from the
        of  $<\!p^{-2}\!>$ and $<\!p^{-1}\!>$ values which,  as
        pointed out  above, sample the interior region of the EMD. The higher the
        momentum density near the origin the greater are the values of these
        moments.  That these moments have largest values in  the SIC formalism
         may be qualitatively explained as follows:
         The region near the origin in the momentum space by Fourier- reciprocity,
         corresponds to   the asymptotic region in the position space.
         In the density functional theory (DFT) the asymptotic decay of
         the co-ordinate space electron density goes as 
         $\sim exp(-2 \sqrt{2\vert\epsilon_{max}\vert} ~r)$, 
where $\epsilon_{max}$ is the
         eigenvalue of the highest occupied orbital\cite{LPS}. The highest
         occupied orbital energy eigenvalues for these systems in the SIC formalism  are
         smaller in magnitude compared to their HF and HS counterparts. 
         Consequently, the coordinate-space electron density decays 
slowly in SIC than in the HS and HF theories, leading therefore to 
higher values of $<\!p^{-2}\!>$
         and $<\!p^{-1}\!>$ moments.  
               The larger values of the HF profiles than HS profiles
         near $q=0$  can also be explained similarly.
           In the HF theory (unlike in DFT), {\em all} the orbitals decay 
with the same
           exponent $(\sim exp(-\sqrt{2\vert\epsilon_{HF}^{max}\vert} ~ r))$
           asymptotically $(\epsilon_{HF}^{max}$, here is the highest occupied HF
           orbital energy eigenvalue)\cite{Handy} which   by reciprocity, reflects in the
          slower decay of $\gp$ in the small $\vert\vp\vert$, resulting therefore 
          in  slightly larger values of  $<\!1/p^2\!>$,  $<\!1/p\!>$ moments
           and $J_{HF}(q)$ (near $q=0$) as compared to their HS
           counterparts.
              Amongst the various moments,   the agreement among 
    these theories is the best for the $<\!p^2\!>$ moments. The HF and HS values
    of this moment   are very close,  agreeing  up to four
    significant figures in case of Ne and Ar, and up to three significant
    figures for Kr.  
        This agreement is however not surprising since this moment is 
    essentially the double of the negative of the total energy (by the virial
 theorem) and the
    HS theory  is known to produce the total energies which are
    practically equivalent to their HF counterparts.
        Further,  the $<\!p\!>$ and $<\!p^3\!> $ moments are also found to be 
    comparable in the three theories. The $<\!p^4\!>$ moments within HS
   agree with their corresponding HF and/or SIC values.
%It  is difficult to reason out this observed discrepancy.
       Thus, the HS theory with its local
    prescription for the exchange potential seen to mimic the
    Hartree-Fock  formalism reasonably well in the momentum space.
  Our study on the detailed structure of the radially projected first order
       reduced density matrix\cite{SRG} also supports this similarity between 
      the HS and HF density matrices: 
       striking similarities are observed in the structure(contours)
       of reduced first order density matrix in the momentum space.

        The  HS theory also  offers a simpler description of
the bound excited  states in comparison with the HF description. This is
 because the HS formalism is not based on 
the variational principle, but rather on the physical effect of the 
Pauli and Coulomb correlations; which has prompted us to compute
the CPs for the excited states of helium. 
%
% Addition: revised 2
%
 Of course, the numerical HF approach is also perfectly suitable for excited
states with the orbitals identified from the number of radial nodes $(=
n-l-1)$ of a given radial part $R_{nl}(r)$ of the orbital. However, the
appeal of the HS approach is that it is simple to implement than the numerical
HF scheme, yielding results that are practically equivalent to the latter.
% Addition: revised 2 over
%
The CP's calculated for
        various excited states of Helium are presented in Table IX.
        The excited states of helium atom will have diffused electron
        distribution  in the position space and also will have higher
        total energy. Consequently, the excited state CP's  will be more
        compact or localized in the small $q-$ region, as is evident from Table
       IX. 

   One naively expects that the  HS Compton profiles may be 
 improved by adding an accurate local correlation  to its effective 
potential. It is observed, however,  that
    addition of an {\em ad-hoc} correlation (such as the Gunnarsson-Lundqvist
     \cite{Gun} or Ceperly-Alder \cite{PZ} prescriptions)
   to the effective HS potential results in an undesirable  lowering of the
   peak-profile.
%
% Addition: revised 2
%
    If the correlation  is added at right the level of exchange-correlation
hole ($\rho_x$ replaced by $\rho_{xc}$ in  Eq.[\ref{Ex:eq}] ) and {\em then} 
the HS computations be performed self-consistently,
 an improvement over the ``exchange only'' $J(q)$ is expected.
On the other hand, though the work of Holas and March \cite{HM} 
as pointed above prescribes an inclusion of the kinetic piece of correlation 
in the KS context  it is not known how one actually implements
their  scheme in practice. These studies of course, go beyond the scope of the
present ``exchange-only'' theme.
%
% Addition: revised 2 over
%

             In this  paper, we have carried out a comparative study 
of the momentum space properties of atoms viz. Compton profiles and 
various expectation values calculated within the ``work formalism'' of 
Harbola and Sahni, the Hartree-Fock theory as well as  the 
Self-Interaction-corrected Local Density Approximation  theory.
The Compton profiles for various excited states of the Helium
atoms are also presented within the work formalism. The present 
work demonstrates that the Harbola-Sahni work formalism which 
in position space closely
follows the HF theory also seen to do so in the momentum space.

          RRZ gratefully acknowledges the financial assistance from
         CSIR, New Delhi. 
         RKP wishes to thank the University Grants Commission, New Delhi, for
financial support.
The authors acknowledge Center for Network 
Computing, Pune University, for computer time.

\narrowtext

\begin{table}

%==========================================================================
\caption{                         
 Spherically averaged Compton profile, $J_{sph}(q)$ for Helium within
the three  ``exchange only '' theories compared with their experimental
counterpart. (Hartree $a.u.$ used throughout)}

\begin{tabular}{ccccc}
%-----------------------------------------------------------------------
      q   &    HF     &    HS    &   SIC          &      Expt.$^a$  \\
%--------------------------------------------------------------------
\tableline
   0.0    &   1.070   &   1.070  &  1.070      &         1.071{$\pm$}1.5\%  \\
   0.2    &   1.017   &   1.017  &  1.017      &         1.019  \\
   0.6    &   0.700   &   0.700  &  0.700      &         0.705   \\
   1.0    &   0.382   &   0.382  &  0.382      &         0.388  \\
   1.5    &   0.160   &   0.160  &  0.160      &          --  \\
   2.0    &   0.068   &   0.068  &  0.068      &         0.069  \\
   2.5    &   0.031   &   0.031  &  0.031      &         0.030{$\pm$}15\%  \\
   3.0    &   0.015   &   0.015  &  0.015      &         0.013  \\
%=======================================================================  \\
\end{tabular}

     $^a$  Ref. \cite{ER}
\end{table}

\begin{table}

%==========================================================================
\caption{  $J_{sph}(q)$ for                   Neon }

\begin{tabular}{cccccc}
%------------------------------------------&---------&------------------
      q   &    HF    &     HS    &   SIC   & CI$^a$      &  Expt.$^b$  \\
%---------------------&-----------&--------&---------&---------------
\tableline
   0.0    &  2.727   &   2.719   &  2.751  & 2.739   &  2.762    \\
   0.2    &  2.696   &   2.687   &  2.717  & 2.707   &  2.738    \\
   0.4    &  2.593   &   2.585   &  2.608  & 2.602   &  2.630    \\
   0.6    &  2.413   &   2.406   &  2.418  & 2.4159  &  2.427    \\
   0.8    &  2.168   &   2.162   &  2.163  & 2.1645  &  2.162    \\
   1.0    &  1.889   &   1.885   &  1.875  & 1.880   &  1.859    \\
   1.5    &  1.228   &   1.228   &  1.211  &  --    &   --       \\
   2.0    &  0.771   &   0.774   &  0.764  & 0.768  &   0.765    \\
   2.5    &  0.501   &   0.506   &  0.501  & --     &   0.501    \\
   3.0    &  0.346   &   0.350   &  0.349  & 0.348  &  0.359    \\
   3.5    &  0.253   &   0.256   &  0.256  &  --    &  0.277    \\
   4.0    &  0.194   &   0.196   &  0.197  &  0.196 &  0.210    \\
   5.0    &  0.125   &   0.125   &  0.126  &  0.126 &  0.126    \\
&
%==========================================================================
\end{tabular}

  $^a$ Ref. \cite{CP_CI}

  $^b$ Ref. \cite{Tong_L}
\end{table}

\begin{table}

%==========================================================================
\caption{  $J_{sph}(q)$ for           Argon         }

\begin{tabular}{ccccc}
%-----------------------------------------------------------------------
      q   &    HF     &    HS     &     SIC   &         Expt.$^a$    \\
\tableline
%-----------------------------------------------&-----------------------
   0.0    &  5.064    &   5.040   &    5.093  &  5.058    \\
   0.2    &  4.963    &   4.941   &    4.991  &  4.917    \\
   0.4    &  4.619    &   4.605   &    4.638  &  4.526    \\
   0.6    &  4.035    &   4.029   &    4.033  &  3.960    \\
   0.8    &  3.333    &   3.331   &    3.312  &  3.319    \\
   1.0    &  2.661    &   2.664   &    2.636  &  2.697{$\pm$}1\%    \\
   1.5    &  1.546    &   1.557   &    1.540  &  --    \\
   2.0    &  1.084    &   1.090   &    1.086  &  1.129    \\
   2.5    &  0.874    &   0.876   &    0.875  &  0.904    \\
   3.0    &  0.736    &   0.736   &    0.737  &  0.744    \\
   3.5    &  0.622    &   0.621   &    0.620  &  0.634    \\
   4.0    &  0.520    &   0.519   &    0.520  &  0.534{$\pm$}2.5\%    \\
   4.5    &  0.433    &   0.432   &    0.432  &  --    \\
   5.0    &  0.359    &   0.359   &    0.359  &  0.366    \\
  10.0    &  0.075    &   0.076   &    0.076  &  0.078{$\pm$}10\%    \\
  15.0    &  0.025    &   0.025   &    0.025  &  0.025    \\
%==========&===============================================================
\end{tabular}

     $^a$ Ref. \cite{ER}
\end{table}

\begin{table}

%==========================================================================
\caption{  $J_{sph}(q)$ for     Krypton               }

\begin{tabular}{ccccc}
%-----------------------------------------------------------------------
      q   &    HF     &     HS   &   SIC  &                 Expt.$^a$  \\
\tableline
%----------&-------------------------------\\% --------------------------
   0.0    &  7.237    &   7.195  &  7.262     &   7.188    \\
   0.2    &  7.095    &   7.060  &  7.122     &   6.988    \\
   0.4    &  6.605    &   6.586  &  6.625     &   6.453    \\
   0.6    &  5.785    &   5.781  &  5.783     &   5.702    \\
   0.8    &  4.855    &   4.863  &  4.841     &   4.883    \\
   1.0    &  4.044    &   4.059  &  4.032     &   4.131{$\pm$}1.7\%    \\
   2.0    &  2.442    &   2.447  &  2.448     &   2.557    \\
   3.0    &  1.858    &   1.857  &  1.854     &   --    \\
   4.0    &  1.327    &   1.324  &  1.319     &   1.350    \\
   5.0    &  0.935    &   0.935  &  0.931     &   0.933{$\pm$}3.5\%    \\
  10.0    &  0.260    &   0.260  &  0.261     &   0.254    \\
  15.0    &  0.105    &   0.105  &  0.105     &   0.099    \\
%==========&============================================================
\end{tabular}
  $^a$ Ref. \cite{ER}
\end{table}

\begin{table}

\caption{
 $<p^n>$ moments for Helium within different ``exchange-only''theories.}
\begin{tabular}{cccc}
 moments  &     HF &    HS   &    SIC \\
\hline
 $<\!p^{-2}\!>$   &     4.0893E+00  & 4.0760E+00   &       4.0902E+00            \\
 $<\!p^{-1}\!>$   &     2.1406E+00  & 2.1409E+00   &       2.1410E+00            \\
 $<\!p>$        &     2.7990E+00  & 2.7990E+00   &       2.7987E+00\\
 $<\!p^{2}\!>$   &     5.7234E+00  & 5.7234E+00   &       5.7138E+00\\
 $<\!p^{3}\!>$   &     1.7991E+01  & 1.7990E+01   &       1.7628E+01\\
 $<\!p^{4}\!>$   &     1.0567E+02  & 1.0549E+02   &       8.7395E+02\\

\end{tabular}

\end{table}

\begin{table}

\caption{
 $<p^n>$ moments for Neon.}
\begin{tabular}{cccc}
 moments  &     HF &    HS   &    SIC \\
\hline
 $<\!p^{-2}\!>$   &     5.4795E+00  & 5.4526E+00   &       5.6349E+00            \\
 $<\!p^{-1}\!>$   &     5.4558E+00  & 5.4387E+00   &       5.5025E+00            \\
 $<\!p>$        &     3.5196E+01  & 3.5269E+01   &       3.5246E+01\\
 $<\!p^{2}\!>$   &     2.5709E+02  & 2.5708E+02   &       2.5771E+02\\
 $<\!p^{3}\!>$   &     3.5843E+03  & 3.5720E+03   &       3.5836E+03\\
 $<\!p^{4}\!>$   &     9.8510E+04  & 9.9418E+04   &       9.9898E+04\\

\end{tabular}

\end{table}

\begin{table}

\caption{
 The $<\!p^n\!>$ momets for Argon.}
\begin{tabular}{cccc}
 moments  &     HF &    HS   &    SIC \\
\hline
 $<\!p^{-2}\!>$   &     1.3107E+01  & 1.2943E+01   &       1.3253E+01    \\
 $<\!p^{-1}\!>$   &     1.0128E+01  & 1.0076E+01   &       1.0187E+01    \\
 $<\!p>$        &     8.8699E+01  & 8.8796E+01   &       8.8793E+01    \\
 $<\!p^{2}\!>$   &     1.0536E+03  & 1.0536E+03   &       1.0538E+03    \\
 $<\!p^{3}\!>$   &     2.4301E+04  & 2.4307E+04   &       2.3997E+04    \\
 $<\!p^{4}\!>$   &     1.1393E+06  & 1.1723E+06   &       5.4391E+06    \\

\end{tabular}

\end{table}

\begin{table}

\caption{
 $<\!p^n\!>$ moments for Krypton.}
\begin{tabular}{cccc}
 moments  &     HF &    HS   &    SIC \\
\hline
 $<\!p^{-2}\!>$   &     1.7478E+01  & 1.7084E+01   &       1.7517E+01  \\
 $<\!p^{-1}\!>$   &     1.4474E+01  & 1.4390E+01   &       1.4524E+01  \\
 $<\!p\!>$        &     2.8141E+02  & 2.8161E+02   &       2.8155E+02  \\
 $<\!p^{2}\!>$   &     5.5041E+03  & 5.5013E+03   &       5.5072E+03  \\
 $<\!p^{3}\!>$   &     2.2628E+05  & 2.2424E+05   &       2.2453E+05  \\
 $<\!p^{4}\!>$   &     1.9852E+08  & 5.0212E+07   &       5.0297E+07  \\
\end{tabular}

\end{table}

% revised   Date: March 16, 1999

\begin{table}
\caption{ $J_{sph}(q)$ for  different states of Helium atom within the
Harbola-Sahni approach.}
\begin{tabular}{cccccc}
 q  &   $1s^2$ &  $1s2s$ & $1s2p$& $1s3p$  &  1s4p     \\
\hline
0.0  &   1.070  &  2.516  & 1.583  &  2.966  &  4.433  \\
0.2  &   1.017  &  1.532  & 1.467  &  1.266  &  0.955  \\
0.4  &   0.879  &  0.592  & 0.949  &  0.517  &  0.423  \\
0.6  &   0.700  &  0.362  & 0.537  &  0.412  &  0.357  \\
0.8  &   0.527  &  0.294  & 0.340  &  0.303  &  0.287  \\
1.0  &   0.382  &  0.237  & 0.239  &  0.228  &  0.223  \\
1.5  &   0.160  &  0.119 &  0.111  &  0.112  &  0.111  \\
2.0  &   0.068  &  0.056 &  0.052  &  0.053  &  0.053  \\
2.5  &   0.031  &  0.029  & 0.025  &  0.025  &  0.025  \\
3.0  &   0.015  &  0.013 &  0.012  &  0.012  &  0.012  \\
\end{tabular}

\end{table}

 \end{document}